# Temporal Consistency Optimization for Alpine Lake Turbulent Flux Observations: A Machine Learning Approach

Zheng Jin[1, 2]

1. Key Laboratory of Land Surface Pattern and Simulation, Institute of Geographic Sciences and Natural Resources Research, Chinese Academy of Sciences, P.R. China
2. University of Chinese Academy of Sciences, Beijing, P.R. China

## Abstract

Aiming to mitigate the temporal inconsistency in eddy covariance (EC) flux observations, an ultra-wide neural network structure is constructed based on the TensorFlow framework, with which the artificial neural networks (ANNs) are more capable of estimating flux intensity via in-situ micrometeorological features. The EC measurements and micrometeorology observations are conducted at the shore of an alpine lake Yamzho Yumco in southern Tibet Plateau (TP). The performance of the ANNs is evaluated via 10-fold cross-validation. As a result, the simulation bias level exhibits minuscule perturbation over different cross-validation subsamples. As an innovative attempt, the micrometeorological features are selected according to their thermodynamic or kinetic information utilization rather than statistical correlations with the flux intensity. The method providing uncertainty mitigation can be extended to other EC flux measurement experiments, especially in harsh regions like TP, where the environmental conditions do not allow more direct observations.

# 1. Introduction

The Tibet Plateau (TP) is a significant heat source to the atmosphere (Ye and Wu 1998) while contains numerous alpine lakes (Ma et al. 2011). Since the environment of the TP has experienced evident impacts of climate change (Yang et al. 2013), the interaction between lakes and the atmosphere is increasingly recognized as an essential scientific issue for water and energy transfer research on the TP. As direct indicators to quantify the lake-atmosphere interaction in mass and energy, the fluxes on lake surface are pivotal to be measured. To meet the demand of flux measurements on lakes or other ecosystems, the application of the eddy covariance (EC) method in measuring energy fluxes between multifarious underlying surfaces and the atmosphere has been widely spread over the past few decades (Baldocchi 2014). The EC method not only calculates flux intensity under a few fluid assumptions but also produces flux data with clear spatial representativity (Schmid 1994). Aiming to derive further cognitions in the high-altitude lake-atmosphere interaction, the EC observations on alpine lakes are initially carried out in the recent years. However, these observations are facing the problem of temporal inconsistency in the turbulent flux data. As previous studies reported, over 60% of flux data were removed in the study of the Nam Co lake (Altitude height: 4715 m) (Wang et al. 2015), about 22% to 30% of flux data were removed in the study of the Ngoring lake (Altitude height: 4274 m) (Li et al. 2015), all flux data from the direction of land (Covers all westwards directions) were removed in the study of the Erhai lake (Altitude height: 2000 m) (Liu et al. 2015). As the primary reason for the eliminations of flux data in the above studies, installing at the lakeshore or somewhere nearby makes the measured flux inevitably contaminated by the contribution of the land surface when the averaged wind direction is pointing at the lake. However, subjected to the strong wind, severe bumping of the floating ice, and high financial cost of equipment maintenance in the depopulated regions, install the instruments in the center of the lake are nearly impossible. Besides, there are strict environmental protection policies in these unexploited areas. As flux data contain gaps with a non-negligible number, these previous studies had employed interpolation methods such as bulk aerodynamic transfer model (B model) (Verburg and Antenucci 2010; Biermann et al. 2014; Wang et al. 2017) and neural network regression (Funahashi 1989; Liu et al. 2015). In the study of the Nam Co lake (Wang et al. 2015), the performance of the B model in flux simulation was straightly evaluated through deriving the mean absolute error (MAE) between the observed and simulated fluxes. As a result, the MAE is 8-9 and 24-29 W m-2 for sensible heat flux and latent heat flux respectively. In the study of the Erhai lake (Liu et al. 2015), an ANN model with 8 neurons in the hidden layers was employed to fill the gaps in flux data, but no validation result about the performance of the ANN model was reported. Therefore, better methods for flux data gap-filling can provide promising mitigation of uncertainty to alpine lake EC observation experiments.

Machine Learning (ML) method is one of the most fast developing technology today, while recent improvements in ML are primarily driven both by innovative algorithms and economic hardware (Jordan and Mitchell 2015). Thus, it is now available to build artificial neural networks (ANNs) with larger size and more complex structures which can prospectively improve their performance in the tasks of interest. The EC measurements produce data with high temporal resolution, meanwhile, are usually conducted with synchronous micrometeorological observations. Since homogeneous observations sustain a long-term operation, favorable materials (EC flux data and micrometeorological data) for ML methods would have been acquired, with appropriate magnitude and dimensionality. Thus, our work takes an attempt to bring newly developed ML algorithms and advanced parallel computing schemes into the EC measurements research. The practical operation of this attempt is to fit the mapping relationship between the EC fluxes and micrometeorological features through more advanced ANNs. In Section 2, firstly, basic information for our observing site is presented. Then in Section 3, the flux data processing of the EC observation is introduced, including footprint analysis and quality control. The tools for establishing and training ANNs are presented. After that, the feature engineering for micrometeorological data is proposed. Then a cross-validation method for evaluating the performance of the ANNs is introduced. In Section 4, the results of footprint analysis are showed firstly. Then the proposed ANN structure and training parameters are presented. Then the temporal patterns of the fluxes which are patched by the simulated data are displayed completely. Most importantly, the results of the cross-validation are presented. In Section 5, conclusions of this study are summarized, and the perspectives of applying ML methods into the EC experiments research are discussed.

## 2. Site and Materials

*a. Site Description*

The Yamzho Yumco lake (Fig.1a) is the largest closed lake in the north foot of the Himalayas (Zhang et al. 2012; Zhe et al. 2017), with an average elevation of 4500 m above sea level, which is regarded as one of the three holy lakes in the Tibetan Buddhism. The flux site (Fig. 1b) is situated near the lakeshore of the Yamzho Yumco, approximately 3 m offshore and over 1 km from the south-eastern lakeshore. Meanwhile, the instruments are fixed on a steel skeleton (Fig. 1c), with a height of 2.1 m above the water surface. As the water level is fluctuating over time, these values of installation information would also change but merely in an acceptable range. To prevent the equipment from been destroyed by floating ice and severe frost, all the instruments would be dismounted in early January and reloaded at the end of March regularly annually.

*b. Instrumentation*

The EC system (Fig. 1c) includes an ultrasonic anemometer (CSAT3, Campbell Scientific®, Inc.) and an open path infrared $H_2O/CO_2$ analyzer (IRGASON, Campbell Scientific®, Inc.). When dismounted before the icing period, the infrared analyzer would be sent to the manufacturing factory to make laboratory level calibration. The micrometeorology observation system includes an air thermo-hygrometer (HMP155A, Vaisala®, Inc.), an infrared thermometer (SI-111, Apogee Instruments®, Inc.) and a radiometer (CNR4, Kipp & Zonen®). The water surface temperature is measured by the infrared thermometer. The upward and downward short-wave and long-wave radiations on the water surface are measured by the radiometer. All the above measurements are synchronized and set to the same sampling frequency. Besides, the systems are powered by solar panels.

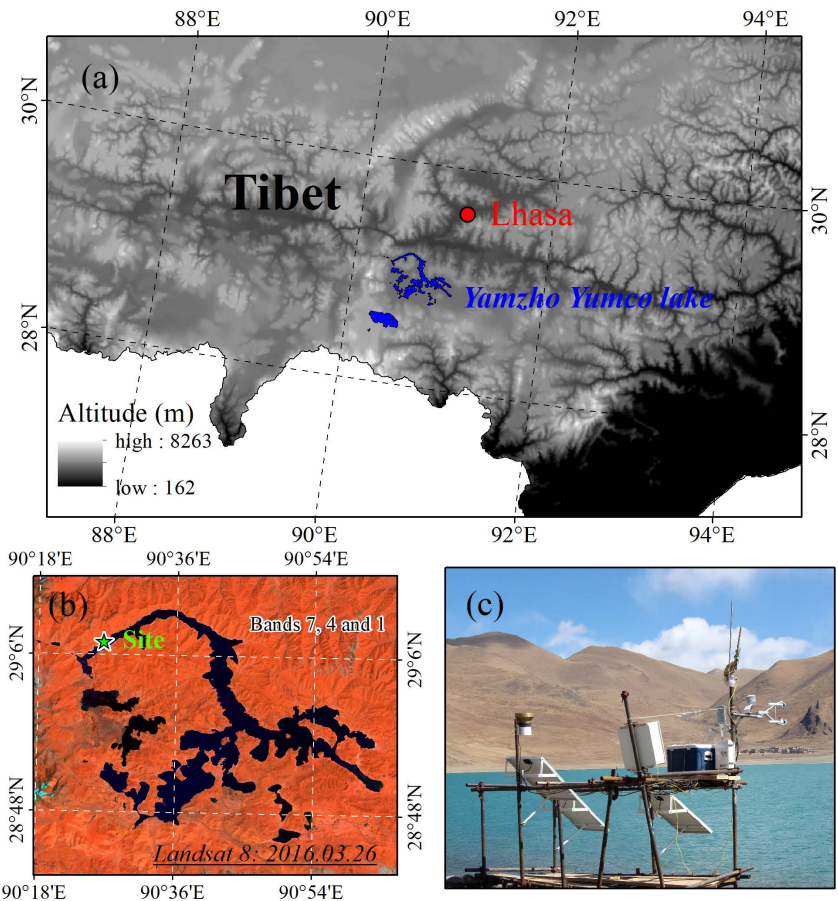

**Fig. 1** (a) Location of the Yamzho Yumco lake (b) Location of the flux site (c) Photograph of the instruments

# 3. Methods

## a. EC Data Processing

The EC turbulent flux calculations and footprint analysis are conducted on the EddyPro® (LI-COR®, Inc.) software platform. The flux intensity is calculated half-hourly with a sampling frequency of 10 Hz, while each value represents the average flux intensity over the past 30 minutes. The corrections for the EC processing include time lag compensation, spike removal, coordinate rotation (Wilczak et al. 2001), WPL density correction (Webb 1982), ultrasonic virtual temperature correction (Dijk et al. 2004; Lee and William 2011) and spectral filtering correction (Fratini et al. 2012). The quality of flux data is classified through a turbulent condition diagnostic method (Göckede et al. 2006). As the result of data quality control, the data points with quality class equal or lower than 2 are eliminated. The footprint analysis is based on the LPDM-B model (Kljun et al. 2004). The EC flux data and micrometeorological data in the period of 3 April to 31 December are employed, both in 2016 and 2017. Another set of data is employed to test the extensibility of the proposed flux simulation method, which covers 4 days before and 4 days after the above period.

## b. Feature Engineering for Micrometeorological Data

The physical controls on half-hourly turbulent flux intensity in the alpine region are studied before (Li et al. 2015; Wang et al. 2017). The previous results demonstrated that several key factors have significant statistical relevance to turbulent flux intensity, such as wind speed and vapor pressure deficit. However, in our research, the selection of inputting feature for the ANNs is not based on the statistical correlation between the feature and flux intensity. The viewpoint on feature selection in this paper is based on the information utilization. Although wind speed, relative humidity and the temperature difference between air and water surface have the most significant statistical correlations to flux intensity, the information about the thermodynamic or kinetic processes in turbulent exchange may not been sufficiently contained in them. For example, the downward short-wave radiation could not directly affect the flux intensity at the time when they are synchronously measured. As a hysteretic effect, when the water is heated by the radiation, it takes time to find the increase in evaporation. Thus, the micrometeorological features which make indirect effects on the flux intensity are also supposed to contain the information about the thermodynamic or kinetic processes in energy transfer. As the ANNs can fit the nonlinear mapping relationship (Specht 1991), while a newlydeveloped regularization algorithm "Dropout" (Srivastava et al. 2014) is presented to refine the "useful" information through the ANNs, it is now feasible to improve the information utilization in ANNs despite the additional information contains higher noise. As a result of the feature engineering, every micrometeorological feature contains 3 values in each temporal point: the 30 minutes average of over the past, at present and in the next. The employed micrometeorological features include air temperature, water surface temperature, averaged wind speed, turbulent kinetic energy, M-O length (Monin and Obukhov 1954), air pressure, vapor pressure, saturated vapor pressure, upward long-wave radiation, upward short-wave radiation, downward long-wave radiation and downward short-wave radiation, counted to 12 in total. Thus, the total dimensionality of the micrometeorological features is 36. All of these features are standardized into the range of 0 to 1.

## c. Establishing ANNs

In this paper, all ANNs are established and trained on the open source ML framework *TensorFlow* (Abadi et al. 2016), with the Python module *Keras* (https://keras.io). The initial structure of the ANNs is densely connected neural networks, without adding on bias terms. The loss function is mean absolute error. The optimizing mode is stochastic gradient descend (Bottou 2010). The regularizing method is Dropout (Srivastava et al. 2014). The activation functions are PReLU (Parametric Rectified Linear Unit) (He et al. 2015) and Tanh (Hyperbolic Tangent) (Fahlman and Lebiere 1990). Considering the magnitude of the input features, the complexity of the fitting algorithms and the structure size of the ANNs, this paper employed the CUDA® (Nvidia®, Inc.) parallel computation scheme with a hardware computation unit GTX-1080® (Nvidia®, Inc.) to make the training time acceptable. The GTX-1080® possesses a peak performance of 9 Tera single-precision floating point operations per second.

*d. Cross-Validation*

The flux simulation performance of the ANNs is evaluated via the ***k***-fold cross-validation (Kohavi 1995), in which the ***k*** is the number of cross-validation subsamples. The determination of ***k*** faces a contradiction between the adequacy of training data and the reliability of cross-validation result. Since the flux data volume of each variable is close to 10,000, to ensure each validation subsample has adequate data in every intensity interval for validation, the number of cross-validation subsamples is determined to be 10. The segmentation of the flux data has to meet 2 requirements to retain the reliability of the cross-validation, one is every subsample should be shaped in a similar density probability distribution to the unsegmented sample, while another is the subsamples and the unsegmented sample should possess a roughly equaled average value. Thus, from the range of -40 to 80 W m$^{-2}$ for sensible heat, -10 to 260 W m$^{-2}$ for latent heat and -1 to 7 mmol s$^{-1}$ m$^{-2}$ for water vapor respectively, these fluxes are divided into 6 equal intervals (Tab. 1). After that, 1/10 data points in each interval are uniformly picked and regrouped into one cross-validation subsample. As there are 10 ANNs which are independently trained through the cross-validation subsamples, the simulations of flux intensity would be the average of the 10 ANNs.

*e. Bias Metrics*

In order to quantitatively analyze the difference between the observed and simulated fluxes, the correlation coefficient ***R*** (Legates and McCabe Jr 1999), the symmetric mean absolute percentage error ***SMAPE*** (Makridakis 1993) and the mean absolute error ***MAE*** (Dawson et al. 2002) are employed, which are defined as:

$$R = \frac{\frac{1}{n}\sum_{i=1}^{n}(S_i - \bar{S}_i)(O_i - \bar{O}_i)}{\sqrt{\frac{1}{n}\sum_{i=1}^{n}(S_i - \bar{S}_i)^2}\sqrt{\frac{1}{n}\sum_{i=1}^{n}(O_i - \bar{O}_i)^2}}, \quad (1)$$

$$SMAPE = \frac{100\%}{n}\sum_{i=1}^{n}\frac{|S_i - O_i|}{(|S_i| + |O_i|)/2}, \quad (2)$$

$$MAE = \frac{\sum_{i=1}^{n}|S_i - O_i|}{n}. \quad (3)$$

Where *n* is the total number of flux data points in one cross-validation subsample, $S_i$ is the ANN simulated fluxes and $O_i$ is the observed fluxes.

**4. Results**

*a. Footprint Analysis*

The footprint plot (Fig. 2) provides the basis of the division for the lake-surface-sourced and land-sourced fluxes. The red spot shows the position of the EC instruments. The mazarine blue scatters represent the 90% cumulative flux contribution point. Based on the result of field trips, regarding the magnetic north as 0° of clockwise direction, flux data of the footprint direction of 70° to 240° are determined as the first part of lake-surface-sourced flux. Then, flux data of the footprint direction of 240° to 270° with the 90% contribution point which located away from the land are determined as the second part of lake-surface-sourced flux. Thus, the rest of the flux data with source areas close to the land or located in land are eliminated totally.

*b. The Ultra-wide ANN Structure*

The result of the ANN training experiments with nearly 1,000 hours of computation unit full loaded runtime demonstrated that the ultra-wide structured ANNs are performing optimally in estimating the turbulent fluxes. Table 2 displays the structure and training parameter settings based on the *Keras* Python module. Not like the popular deep neural networks, there are only 4 layers in the proposed ANN structure. But in the 2 hidden layers, there are 2048 and 1024 neurons respectively, which shaped an ultra-wide ANN structure. The ANNs that are trained to simulate sensible heat, latent heat and water vapor flux share the same set of 10 micrometeorological features. Besides, the 3 fluxes are simulated using the same ANN structure, while the learning rate and the number of iterations is adapted to different data noise levels for the 3 fluxes. The training experiment tested ANNs with $2^4$ to $2^{12}$ neurons in 2 hidden layers, consequently determined the neurons number combination which has the best flux simulation performance. As the number of neurons expands, the time-consuming of the ANN training with the computation unit quickly turns unacceptable. Thus, the ANNs with the number of neurons bigger than $2^{13}$ are not tried.

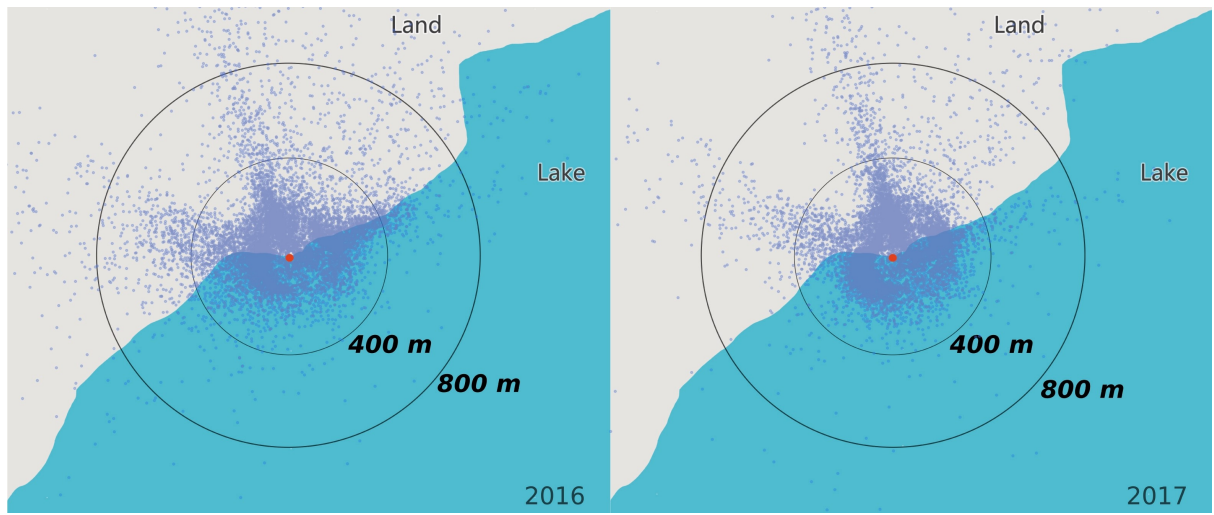

**Fig. 2** Footprint analysis results of 90% cumulative flux contribution points in the observation periods in 2016 and 2017.

*c. The Patching on Flux Data*

Patched with the ANN simulated values, the flux data have acquired higher temporal coverage rates (Tab. 3). After the elimination of the flux data which not principally source from the lake surface and the filtering of quality control, the temporal coverage rates in the observation periods are down to 0.383 or lower. Due to the malfunction of the instruments, data of latent heat flux and water vapor flux in the late October in 2016 turns abnormal. To retain the data quality, the flux data which is observed after September in 2016 is fully eliminated. Since the flux data are generated with significant inconsistency (Fig. 3), the discussions on energy closure or flux variation characteristics of the lake surface are impossible to carry out. Besides, with the data gaps occur without regularity, ordinary regression methods for interpolation would face problems of nonlinearity. There are still data gaps (less than 2%) in the patched fluxes, the reason for the existence of these gaps is the absences of the micrometeorological data at the time.

The diurnal inconsistency patterns in 2016 and 2017 are similar (Fig. 3), in which the large blanks often occur at 00:00 to 09:00, while the coverage rates in 09:00 to 18:00 are higher than other time periods. As the source area is dependent on wind direction and speed, the inconsistent patterns indicate there are significant circulations of the land-lake breeze.

Since flux data have been patched by ML methods, the more complete and clear temporal variation characteristics of lake surface fluxes are revealed (Fig. 3). As for the sensible heat flux in 2016 and 2017, whose intensity shows a steady and significant diurnal variation from April to June, while the diurnal intensity peak often occurs at 10:00 to 16:00. In the late July, the amplitude of diurnal variation declines, while the integral intensity of flux increases. From the late October to the end of November, the high flux intensity period lasts over 30 days, then quickly returns to the level of the early June. The integral flux intensity of sensible heat is greater in 2017 than in 2016, while the integral flux intensity of latent heat is opposite of this.

Then, as for the latent heat flux, the data points in 2016 are 50% less than in 2017. Hence, without the patching of the simulated data, the peak period of October to November in 2016 would not be displayed. In the late November of 2016, the lake experienced a strong evaporation event of nearly a half month, starts and ends in the afternoon of every day, which turns out to be the strongest in the observation periods of 2016 and 2017. The amplitude of temperature variation is smaller than 20°C in the whole observation period both in 2016 and 2017, to this extent, the phase-transition heat of water vapor only changes in a limited ratio of 0.03. Therefore, the variation characteristics of latent heat flux and water vapor flux are significantly similar. In general, the energy release of the lake is a gradually enhancing process, which covers the period of April to October and reaches its peak at the end of October in 2016, while the peak of the observation period in 2017 is postponed to the end of November.

According to the bias expectation of water vapor simulation which derived from the cross-validation, the total evaporation of the *Yamzho Yumco* lake is 740 ± 9 mm and 703 ± 8 mm in the observation period of 2016 and 2017 respectively. In the observation period of 2016, the mean intensity of sensible heat flux, latent heat flux and net radiation is 14.98 ± 0.2, 78.96 ± 0.2 and 126.75 W m$^{-2}$ respectively, while in 2017 is 14.35 ± 0.2, 74.42 ± 0.2 and 139.59 W m$^{-2}$ respectively. Therefore, the energy exchange status of the lake surface is net receiving, if ignored the energy transfer of the internal water flow, the energy received by the lake surface could increase the temperature of the 10 m deep water by 0.1°C a day.

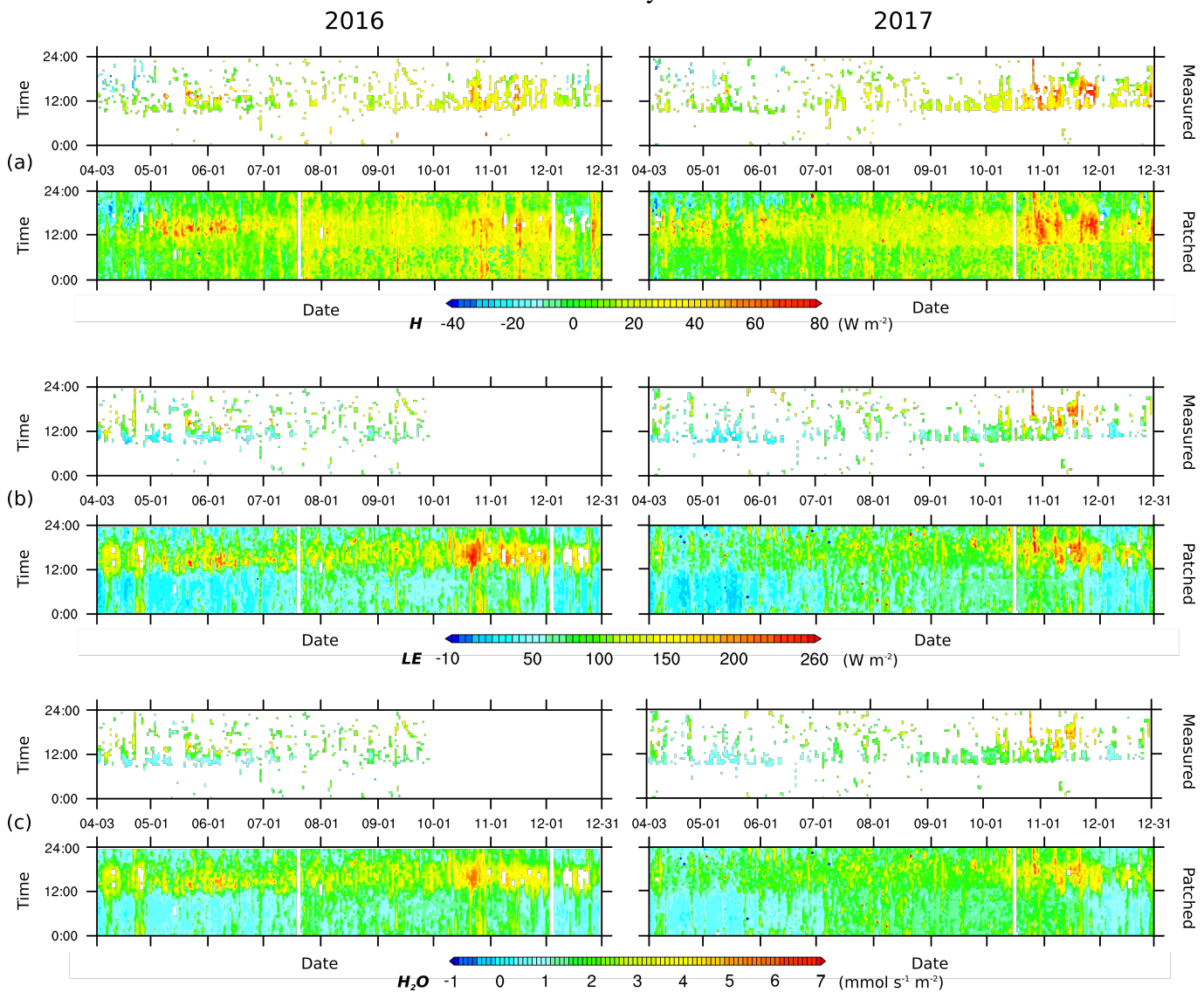

**Fig. 3** Comparison between the unpatched and patched lake surface fluxes: (a) sensible heat (b) latent heat (c) water vapor, with palette resolutions of 5 W m$^{-2}$, 2 W m$^{-2}$ and 0.1 mmol s$^{-1}$ m$^{-2}$ respectively, all included in the period of Apr. 3 to Dec. 31 in 2016 and 2017.

Tab.1 Number of flux data points in each flux intensity interval. Term **H**, **LE** and **H₂O** are the sensible heat flux, latent heat flux and molar water vapor flux with interval lengths of 20 W m$^{-2}$, 45 W m$^{-2}$ and 1.34 mmol s$^{-1}$ m$^{-2}$ respectively.

| Flux intensity intervals | Number of flux data points | | | | | |
|---|---|---|---|---|---|---|
| | 2016 | | | 2017 | | |
| | *H* | *LE* | *H₂O* | *H* | *LE* | *H₂O* |
| Interval 1 | 117 | 431 | 35 | 96 | 537 | 62 |
| Interval 2 | 384 | 1586 | 1870 | 352 | 2090 | 2318 |
| Interval 3 | 2393 | 938 | 1221 | 2232 | 1294 | 1687 |
| Interval 4 | 1609 | 377 | 334 | 1420 | 516 | 507 |
| Interval 5 | 393 | 93 | 54 | 497 | 203 | 157 |
| Interval 6 | 103 | 29 | 13 | 141 | 94 | 35 |

Tab.2 Parameter settings for proposed ANN structure. The averaged learning rate and the averaged iterations are the mean values of the trained ANNs for 10 cross-validation subsamples.

| | Type | Value | | |
|---|---|---|---|---|
| | | *H* | *LE* | *H₂O* |
| Number of neurons | Input layer | 36 | | |
| | Hidden layer 1 | 2048 | | |
| | Hidden layer 2 | 1024 | | |
| | Output layer | 1 | | |
| Activation for layers | Input layer | PReLU | | |
| | Hidden layer 1 | Tanh | | |
| | Hidden layer 2 | PReLU | | |
| | Output layer | Linear | | |
| Dropout rate | | 0.5 | | |
| Averaged learning rate | | 0.0436 | 0.0547 | 0.0542 |
| Learning rate decay rate | | $10^{-6}$ | | |
| Averaged iterations | | 63650 | 91975 | 90725 |

*d.'The 10-fold Cross-Validation*

The performance of the ANNs in estimating flux intensity is evaluated through a 10-fold cross-validation, statistical indicators of the 10 subsamples are presented (Tab. 4). The results show the bias level fluctuates slightly over different cross-validation subsamples, which indicate the ANNs have stable accuracies in flux simulation. As a key performance indicator of the ANNs, the mean value of ***SMAPE*** for ***H***, ***LE*** and ***H$_2$O*** in the 10 subsamples is 31.871%, 20.868% and 20.760% respectively, indicating that the ANNs are performing better for ***LE*** and ***H$_2$O*** than ***H***. Besides, the mean value of ***MAE*** for ***H***, ***LE*** and ***H$_2$O*** in the 10 subsamples is 5.417 W m$^{-2}$, 15.763 W m$^{-2}$ and 0.354 mmol s$^{-1}$ m$^{-2}$ respectively, considering their mean values in total, the simulation performance for ***LE*** and ***H$_2$O*** is also better on ***MAE***.

The differences between AVG-Obv. and AVG-Est. are quite small in all the cross-validation subsamples for H, LE and H2O. As a result of evaluating the simulations on average values, the percentage ratio of error expectations for H, LE and H2O is 2.09%, 1.37%, and 1.83% respectively. Since the biases of the ANN simulated values have a significant symmetry (Fig. 4), the more values taken into the average calculation, the smaller the bias of the mean value would be. But the bias would not decrease infinitely, in this paper, calculating the mean value with 400 or 800 simulated data points makes little difference to the bias.

The comparisons of regression analyses (Fig. 5) show, the results of the subsamples with maximum MAE are significantly similar to those with minimum MAE, indicating the segmentation of training sample has adequate homogeneity, all regression results have passed the 99.9% significance test. A quantile box plot (Fig. 6) shows the mean values of SAPE are smaller than their medians in all cross-validation subsamples.

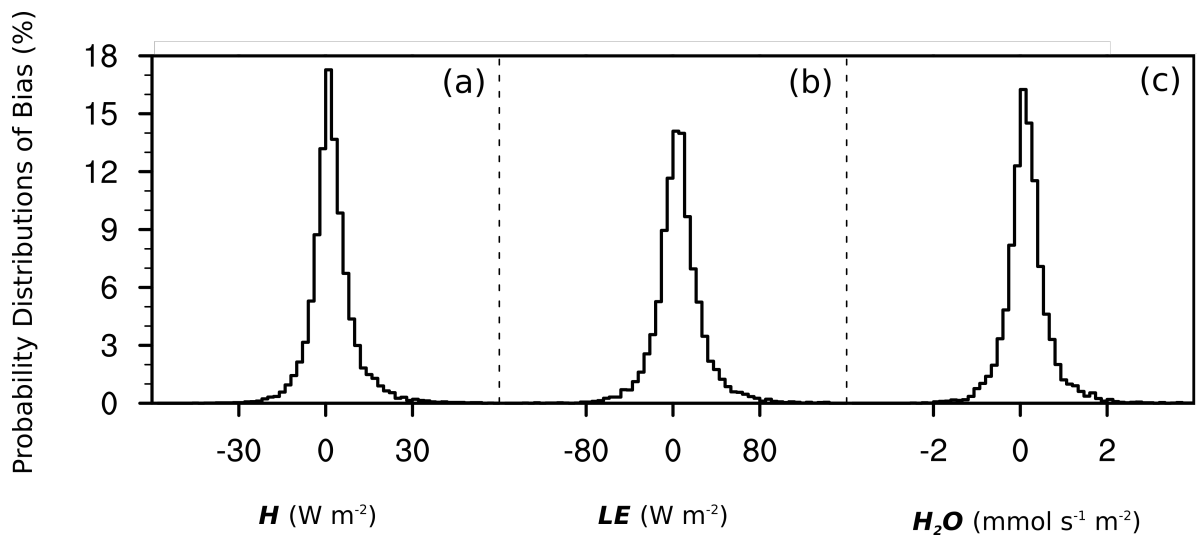

**Fig. 4** Probability distributions of biases in ANN estimated fluxes: (a) sensible heat (b) latent heat (c) molar water vapor.

Tab.3 Temporal coverage rates of lake surface fluxes.

| | Status | Coverage rate | | |
| --- | --- | --- | --- | --- |
| | | H | LE | H$_2$O |
| 2016 | Unpatched | 0.383 | 0.264 | 0.264 |
| | Patched | 0.985 | 0.985 | 0.985 |
| 2017 | Unpatched | 0.367 | 0.365 | 0.365 |
| | Patched | 0.992 | 0.992 | 0.992 |

Tab.4 Results of the 10-fold cross-validation of the ANN simulations. F1 to F10 are the 10 cross-validation subsamples respectively. Term *AVG-Obv.* and *AVG-Est.* are the averaged observations and estimations respectively. The unit of *SMAPE* is %. The units of *MAE*, *AVG-Obv.* and *AVG-Est.* are W m$^{-2}$ for *H* and *LE*, and mmol s$^{-1}$ m$^{-2}$ for *H$_2$O* respectively.

| Statistics | H | | | | | | | | | |
|---|---|---|---|---|---|---|---|---|---|---|
| | F1 | F2 | F3 | F4 | F5 | F6 | F7 | F8 | F9 | F10 |
| **R** | 0.795 | 0.814 | 0.803 | 0.830 | 0.857 | 0.800 | 0.845 | 0.833 | 0.829 | 0.801 |
| **SMAPE** | 32.72 | 32.07 | 31.21 | 31.34 | 31.17 | 33.56 | 31.11 | 30.62 | 31.46 | 33.45 |
| **MAE** | 5.432 | 5.437 | 5.556 | 5.159 | 5.504 | 5.615 | 5.150 | 5.220 | 5.349 | 5.756 |
| **AVG-Obv.** | 18.72 | 18.69 | 19.00 | 18.85 | 19.02 | 18.67 | 18.75 | 18.77 | 19.04 | 19.00 |
| **AVG-Est.** | 18.31 | 19.00 | 18.80 | 18.54 | 19.65 | 18.30 | 18.56 | 18.78 | 18.53 | 17.99 |
| | LE | | | | | | | | | |
| | F1 | F2 | F3 | F4 | F5 | F6 | F7 | F8 | F9 | F10 |
| **R** | 0.816 | 0.766 | 0.782 | 0.765 | 0.812 | 0.797 | 0.813 | 0.778 | 0.779 | 0.770 |
| **SMAPE** | 20.65 | 20.49 | 20.68 | 20.96 | 20.48 | 20.55 | 20.92 | 21.18 | 21.03 | 21.74 |
| **MAE** | 15.23 | 15.69 | 15.55 | 15.76 | 15.83 | 15.48 | 15.75 | 15.83 | 16.13 | 16.38 |
| **AVG-Obv.** | 80.44 | 80.74 | 81.36 | 81.29 | 82.40 | 81.68 | 81.66 | 81.82 | 82.18 | 81.47 |
| **AVG-Est.** | 81.02 | 79.25 | 81.19 | 78.24 | 82.13 | 79.82 | 82.26 | 82.68 | 81.15 | 82.73 |
| | H$_2$O | | | | | | | | | |
| | F1 | F2 | F3 | F4 | F5 | F6 | F7 | F8 | F9 | F10 |
| **R** | 0.773 | 0.779 | 0.762 | 0.763 | 0.792 | 0.773 | 0.822 | 0.809 | 0.774 | 0.767 |
| **SMAPE** | 20.24 | 21.08 | 21.80 | 20.70 | 19.92 | 19.62 | 20.39 | 20.77 | 21.61 | 21.47 |
| **MAE** | 0.342 | 0.353 | 0.373 | 0.360 | 0.345 | 0.328 | 0.348 | 0.351 | 0.370 | 0.370 |
| **AVG-Obv.** | 1.847 | 1.824 | 1.823 | 1.862 | 1.853 | 1.830 | 1.847 | 1.846 | 1.851 | 1.849 |
| **AVG-Est.** | 1.822 | 1.802 | 1.773 | 1.808 | 1.809 | 1.845 | 1.902 | 1.823 | 1.827 | 1.823 |

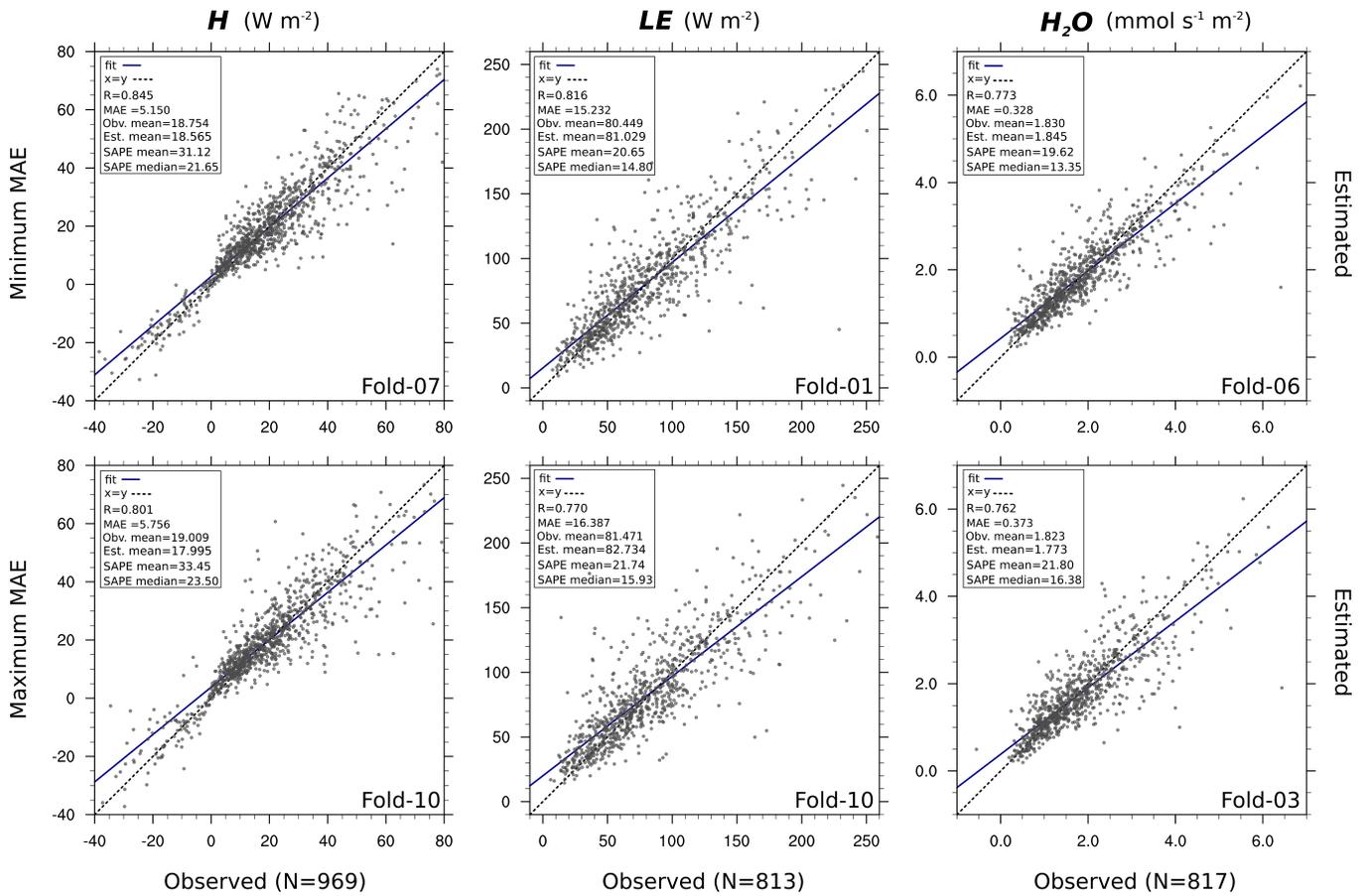

**Fig. 5** Comparisons of the regression analyses results between the subsamples with minimum and maximum *MAE* in the 10-fold cross-validation. Term SAPE is the symmetric absolute percentage error.

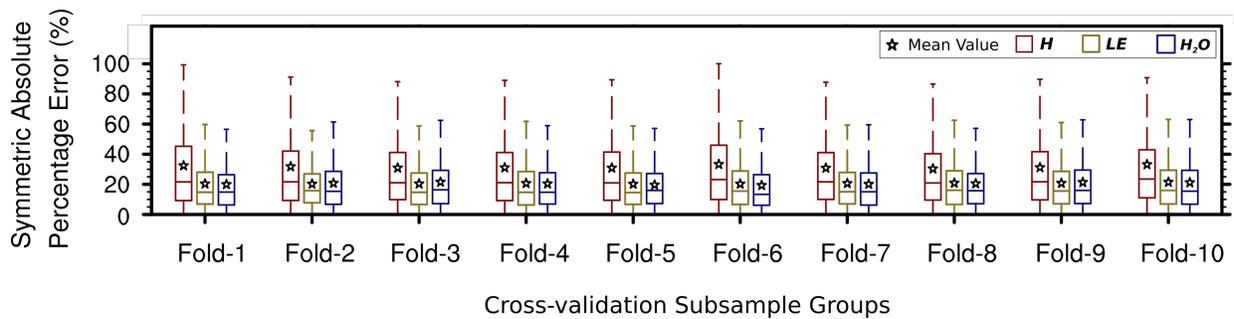

**Fig. 6** Quantile boxes of the symmetric absolute percentage error in the 10-fold cross-validation.

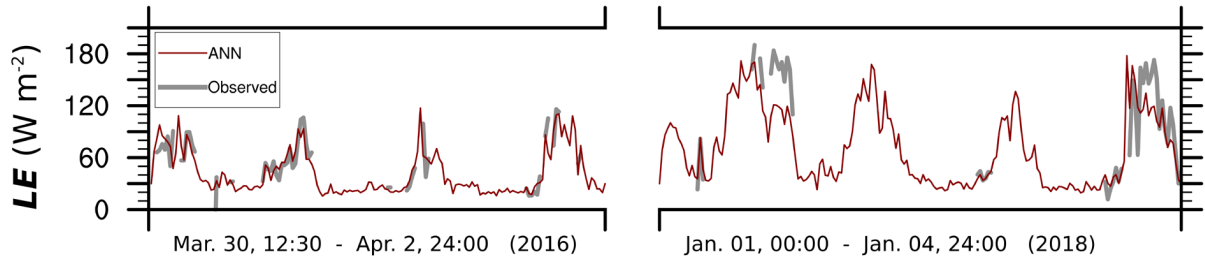

**Fig. 7** Comparison between the ANN estimated and observed latent heat flux.

Especially for sensible heat flux, in which the differences between medians and means are greater than those in the latent heat flux or water vapor flux. Hence, the **SMAPE** is significantly affected by the extreme large values. Finally, the ANN simulated latent heat flux is compared with a bunch of observations which did not used for ANN training (Fig. 7). Like flux data of other periods, the latent heat flux observations in the periods of 30 March to 2 April in 2016 and 1 to 4 January in 2017 are discontinuous, but enough for an intuitive comparison. As Fig. 7 shows, the ANN simulated latent heat flux agrees well with the observations in the intensity interval of 30 to 120 W m$^{-2}$, while claims significant bias when the flux intensity is greater than 120 W m$^{-2}$ or less than 30 W m$^{-2}$. Although the bias is larger within a certain intensity range, the simulated flux and the observations are highly consistent in trends.

## 5. Conclusion and Discussion

In this study, a ML method for the gap-filling in the EC observation was presented. With the proposed ML method, our work revealed high temporal resolution variation characteristics of heat and water turbulent fluxes on the lake surface of the *Yamzho Yumco* for the first time. In the selection of micrometeorological features, this paper avoided the common operation which relies on the statistical correlation between individual features and target variables. Instead, the adequate information utilization of thermodynamic and kinetic background forcing field was primarily taken into consideration. To complete the training computation for the massive structured ANNs, this paper employed the parallel computing scheme CUDA®. Promising results of flux hindcasting indicate that there are broad prospects in applying big data theories into the atmosphere boundary layer observations like EC flux measurement. When validating the simulation performance of the ANNs, this paper found the flux simulation works better in the intensity intervals with larger data volume, demonstrating the flux simulation performance has the potential to improve with the increase of data volume within a certain range. Regarding the ANN training and parameter adjusting as data mining procedures, the information mining in the mapping relationship between the 2-dimensional features and 1-dimensional fluxes has achieved promising results, indicating that adding the spatial dimensionality of features could also possibly improve the performance of the ANNs in flux simulation due to the increase of effective information.

*Acknowledgments.* The author would like to thank Tsering Mima of the *Baidi* hydrological station of Tibet, Dr. Meng Zhe of Tianjin normal university, and Chao Yang and Weiwei Kong of Beijing *Tiannuojiye* Technology Co. Ltd. for their assistance in installation and maintenance of field instruments. This work was supported by the National Natural Science Foundation of China, No. 41471064.

Schmid, H. P., 1994: Source areas for scalars and scalar fluxes. *Boundary-Layer Meteorology*, **67 (3)**, 293-318, https://doi.org/ 10.1007/BF00713146.

Specht, D. F., 1991: A general regression neural network. *IEEE transactions on neural networks*, **2 (6)**, 568-576, https://doi.org/10.1109/72.97934.

Srivastava, N., Hinton, G., Krizhevsky, A., Sutskever, I., and Salakhutdinov, R., 2014: Dropout: a simple way to prevent neural networks from overfitting. *The Journal of Machine Learning Research*, **15 (1)**, 1929-1958.

Van Dijk, A., Moene, A. F., and De Bruin, H. A. R., 2004: The principles of surface flux physics: theory, practice and description of the ECPACK library. *Internal Rep*, *1*, **99**, http://www.met.wau.nl/projects/jep.

Verburg, P., and Antenucci, J. P., 2010: Persistent unstable atmospheric boundary layer enhances sensible and latent heat loss in a tropical great lake: Lake Tanganyika. *Journal of Geophysical Research: Atmospheres*, **115 (D11)**, https://doi.org/10.1029/2009JD012839.

Wang, B., Ma, Y., Chen, X., Ma, W., Su, Z., and Menenti, M., 2015: Observation and simulation of lake-air heat and water transfer processes in a high-altitude shallow lake on the Tibetan Plateau. *Journal of Geophysical Research: Atmospheres*, **120 (24)**, 12327-12344, https://doi.org/10.1002/2015JD023863.

Wang, B., Ma, Y., Ma, W., and Su, Z., 2017: Physical controls on half-hourly, daily, and monthly turbulent flux and energy budget over a high-altitude small lake on the Tibetan Plateau. *Journal of Geophysical Research: Atmospheres*, **122 (4)**, 2289-2303, https://doi.org/10.1002/2016JD026109.
**16 of 17**